\newcommand{\lapprox}{<\atop{\sim}}  % math mode only!
\newcommand{\gapprox}{>\atop{\sim}}  % math mode only!
\newcommand{\aox}{\ifmmode{\alpha_{ox}} \else $\alpha_{ox}$\fi}
\newcommand{\ax}{\ifmmode{\alpha_x} \else $\alpha_x$\fi}
\newcommand{\aE}{\ifmmode{\alpha_E} \else $\alpha_E$\fi}
\newcommand{\loglopt}{\ifmmode{\rm log}~l_{opt} \else log$~l_{opt}$\fi}
\newcommand{\loglx}  {\ifmmode{\rm log}~l_x \else log$~l_x$\fi}
\newcommand{\lopt}{\ifmmode l_{opt} \else $~l_{opt}$\fi}
\newcommand{\lx}  {\ifmmode l_x \else $~l_x$\fi}
\newcommand{\fcgs}{\ifmmode erg~cm^{-2}~s^{-1[B}\else
erg~cm$^{-2}$~s$^{-1}$\fi}
\newcommand{\lcgs}{\ifmmode erg~~s^{-1[B}\else erg~s$^{-1}$\fi}
\newcommand{\fnucgs}{\ifmmode erg~cm^{-2}~s^{-1}~Hz^{-1}\else
erg~cm$^{-2}$~s$^{-1}$~Hz$^{-1}$\fi}
\newcommand{\lnucgs}{\ifmmode erg~s^{-1}~Hz^{-1}\else
erg~s$^{-1}$~Hz$^{-1}$\fi}
\newcommand{\mone}{\ifmmode ^{-1}\else$^{-1}$\fi}
\newcommand{\mtwo}{\ifmmode ^{-2}\else$^{-2}$\fi}
\newcommand{\arcsec}{\ifmmode ^{\prime\prime}\else$^{\prime\prime}$\fi}
\newcommand{\ros}{{\em ROSAT{\rm}}}
\newcommand{\ein}{{\em Einstein{\rm}}}
\newcommand{\nh}{\ifmmode{\rm N_{H}} \else N$_{H}$\fi}
\newcommand{\nhgal}{\ifmmode{ N_{H}^{Gal}} \else N$_{H}^{Gal}$\fi}
\newcommand{\nhintr}{\ifmmode{ N_{H}^{intr}} \else N$_{H}^{intr}$\fi}
\newcommand{\nhtot}{\ifmmode{ N_{H}^{tot}} \else N$_{H}^{tot}$\fi}

\documentstyle[aaspptwo]{article}
\received{9/12/95}
\accepted{12/1/95}
\tighten

\lefthead{Green \& Mathur }
\righthead{BALQSOs Observed by the ROSAT PSPC}

\begin{document}

\title{BROAD ABSORPTION LINE QSOS \\ OBSERVED BY THE {\em ROSAT} PSPC}

\author{Paul J. Green\altaffilmark{1,2} and
Smita Mathur}

\affil{Harvard-Smithsonian Center for
Astrophysics, 60 Garden St., Cambridge, MA 02138}
\altaffiltext{1}{Hubble Fellow}
\altaffiltext{2}{pgreen@cfa.harvard.edu}

\begin{abstract}

Recent results from the \ros\, All-Sky Survey (RASS) have shown that
broad absorption line (BAL) QSOs are either highly absorbed or
underluminous in the soft X-ray bandpass.  Here, we extend this work
by analyzing all known {\em bona fide} BALQSOs observed within the
inner 20$'$ of the \ros\, PSPC.  This sample includes both targeted and
serendipitous exposures ranging from 8 to 75 ksec.  Despite these deep
exposures, most of the BALQSOs are undetected, and have unusually weak
X-ray emission, as evidenced by large optical-to-X-ray slopes \aox.
Large values of \aox\, ($\gapprox$1.8) may prove to be a defining
characteristic of
BALQSOs. We predict that samples of QSO candidates with large \aox\,
will yield a higher percentage of BALQSOs, particularly at low
redshift. As a corollary, X-ray-selected QSO samples should yield
fewer BALQSOs.

The optical/UV emission line spectra of BAL and non-BAL QSOs are quite
similar, suggesting that their {\em intrinsic} spectral energy
distributions are similar as well.  Absorption thus seems the likely
reason for the X-ray quiet nature of BALQSOs.  To constrain the total
absorbing column of the BAL clouds, we compare our measured soft X-ray
fluxes or upper limits to those expected from normal radio quiet QSOs
of comparable optical continuum magnitude and redshift.  From
sensitive X-ray observations, we derive column densities of
$\gapprox$2$\times10^{22}$ cm\mtwo\, for intrinsic cold absorbers of solar
metallicity.  These new results suggest columns {\em at least} an
order of magnitude larger than the columns previously estimated from
optical/UV spectra alone.

\end{abstract}


\section{INTRODUCTION}
\label{intro}

  About 10 - 15\% of optically-selected QSOs have optical/UV spectra
showing deep, wide absorption troughs, displaced to the blue of their
corresponding emission lines.  Occasionally these broad absorption
lines (BALs), seen in the high ionization transitions of C\,IV,
Si\,IV, N\,V, and O\,IV, are also visible in low ionization species,
like Mg\,II and Al\,III.  The BALs are found only among the
radio-quiet QSO population (Stocke et al. 1992), and are suspected to
result from a line of sight passing through highly ionized, high
column density (log$N_H\sim 20$ cm\mtwo) absorbers, flowing outward
from the nuclear region at speeds up to 0.1 - 0.2c.

  For years, BALQSOs were debated to be either a different class of
QSO, or normal radio-quiet (RQ) QSOs viewed at an orientation
traversing a BAL cloud (thus implying that the BAL ``covering factor''
is 10 - 15\%).  More recently, the low BAL cloud covering factors and
absence of emission lines at the high velocities observed in BALQSOs
(Hamann, Korista, \& Morris 1993), along with the similarity of
emission-line and continuum properties of BAL and non-BALQSOs (Weymann
et al. 1991, hereafter WMFH) suggest that orientation is indeed the
cause of the BAL phenomenon, i.e. that {\em all} radio-quiet QSOs have
BAL clouds.  Thus, BALQSOs provide a unique probe of conditions near the
nucleus of most QSOs.  Still, the geometry, covering factor,
temperature, density, metallicity and ionization parameter of the
absorbing clouds are poorly understood from optical/UV (OUV)
absorption line studies (Lanzetta et al. 1991). This is because only a
few lines are measured (e.g., OVI$\lambda 1035$, CIV$\lambda1549$,
NV$\lambda1240$), yielding column densities of a few ions, but little
information on the ionization state. Since the X-ray absorption is
relatively insensitive to ionization and depletion (Morrison \&
McCammon 1983), X-ray spectra give a total column density.  Techniques
have recently been developed by Mathur (1994) and Mathur et al. (1994)
that simultaneously exploit UV and X-ray spectra to constrain the
allowed ranges of the BAL cloud conditions.  An application of these
techniques to BALQSOs may eventually provide stronger constraints on
BAL clouds, but until recently, the X-ray properties of BALQSOs as a
class were unknown since few reports of X-ray detections of BALQSOs
exist.  Furthermore, no homogeneous sample of BALQSOs
with available X-ray data existed.  However, Green et al. (1995)
recently investigated a sample of 36 BALQSOs uniformly selected from
908 QSOs observed during the ROSAT All-Sky Survey in the Large Bright
Quasar Survey (LBQS; see Hewett, Foltz \& Chaffee 1995 and references
therein).  Using a Monte Carlo technique, they found that the average
number of PSPC counts for BALQSOs (for an effective total exposure
time of about 21\,ksec) was lower than the corresponding number for
carefully chosen comparison samples at least 99.5\% of the time.  This
indicates that BALQSOs are (1) either highly absorbed, (2)
intrinsically underluminous in X-rays, or both.  The great similarity
of OUV {\em emission} lines in BAL and non-BALQSO samples leads
us to suspect that their ionizing continua are similar.  This,
together with known OUV absorption, make strong absorption seem
the likely explanation for the paucity of BALQSO detections in the
soft X-ray bandpass.

In this paper, to better constrain the absorption, we
study deep, pointed observations of BALQSOs by the \ros\, PSPC
that are available in the public data archive. We choose a sample of
high-quality, deep soft X-ray images that provides several detections,
as well as much more sensitive flux upper limits for about a dozen
BALQSOs.  In every case, we estimate the intrinsic cold column
density, \nhintr, necessary to absorb the X-ray emission expected from
an average radio-quiet (RQ) QSO.  We also briefly consider the effects
of metallicity and ionization on these estimates.

\section{Sample}
\label{sample}

  We first generated a list of optical BAL positions from Hewitt
\& Burbidge 1993, Junkarrinen, Hewitt \& Burbidge (1991), and from an
updated list kindly provided by Tom Barlow (private communication).
{}From this list, we removed objects that are not listed definitively
as BALQSOs based on analysis of OUV spectra by all these authors.  We
then cross-correlated our list with the list of \ros\, PSPC archived
pointings (version May 18, 1995).  A matching tolerance of 20$'$ from
each optical BAL position includes both serendipitous and targeted
observations, but avoids the ribs of the PSPC.  This inner area of the
PSPC also boasts the narrowest point source response function, thereby
optimizing signal-to-noise ratio (SNR) and decreasing source
confusion.  Lastly, we removed from the sample one QSO (0042-267)
lying close to a
rib, and two QSOs (0104+318 and 0413-116) lying $<1'$ from known
extended X-ray sources (NGC\,383 and Abell\,483, respectively).  The
resulting sample of 12 BALQSOs is listed in Table\,1. Five of these
(with off-axis distances $>1'$ in Table\,1) were observed
serendipitously or as part of a general field, while the rest were
targeted observations of the QSO.  The resulting sample is neither
complete nor uniformly-selected, but it does minimize contamination
{}from normal QSOs or from QSOs with narrow-line/intervening absorption
systems.

\section{Data and Analysis}
\label{data}

Table\,1 shows $B$ magnitudes, redshift and Galactic column density
\nhgal\, as well as the raw X-ray data for each BALQSO in our sample.
We derived net PSPC source counts using circular apertures with radii
$R_{ap}$ (in arcseconds) listed in Table\,1.  Small radii were
sometimes required to avoid nearby unrelated X-ray sources.  To
compensate for the exclusion of source flux from small apertures, we
apply an aperture correction factor to source counts (taken from
Fig.\,11 of Brinkmann et al. 1994), for both detections and upper
limits.  Background counts are shown normalized to the source aperture
area, although we use much larger areas for background estimation, and
exclude any overlapping sources.  Upper limits to the source counts
are set at $3\sigma$.

For comparison of the observed count rate or its upper limit to that
expected from a typical RQ QSO, we first derived an optical luminosity
derived from the observed magnitude. Optical luminosities are
calculated using the magnitudes listed in Table\,1, with $H_0=50$ km
s\mone \,Mpc\mone \,and $q_0=0.5$, and specific optical zeropoint
from Marshall et\,al. (1984).  We include a reddening correction of
$E_{B-V}={\rm max}[0,(-0.055 + 1.987\times10^{-22}\nhgal]$ and
$A_V=3E_{B-V}$. We then predict an intrinsic soft X-ray flux using the
slope \aox\, of a hypothetical power law connecting rest-frame
2500\,\AA\, and 2\,keV, where $\aox\, = 0.384\,{\rm log} (\frac{
\lopt}{\lx })$.  Objects with large \aox\, thus have stronger
optical emission relative to X-ray.  We assume $\aox=1.6$, the mean
for RQ QSOs in the \ros\, PSPC (Green et al. 1995).  Using our
predicted soft X-ray flux, we then used PIMMS
(the Portable Interactive Multi-Mission Simulator; Mukai 1993)
to derive the expected PSPC count rate corresponding to this flux,
assuming a power-law model
with $\alpha_E=1.5$ (the mean PSPC $\alpha_E$ for radio-quiet quasars;
Laor et al. 1994) and the Galactic column density. If the expected
PSPC count rate from PIMMS was greater than observed (as is the case
for all BALQSOs in our sample), we increased the column density of the
absorber in the spectral model until the observed count rate or upper
limit was reached.  This results in a {\em lower} limit to the
`observed' column, on the assumption that the absorber is cold gas
with solar metallicity.  After deriving the ($z=0$) additional column
required to match the PSPC count rate or upper limit for the BALQSOs
in our sample, we then used XSPEC (an X-ray spectral fitting package;
Arnaud 1995) to calculate the column density of
the absorber at the source redshift $z_{em}$.  Using the January 1993
PSPC response matrix, we first simulated a PSPC spectrum of a QSO with
$\aE=1.5$.  This slope is used together with the corresponding total
($z=0$) absorption and model normalization found via PIMMS.  We then
refit the simulated spectrum with a 5 parameter model: (1) Galactic
column density (fixed at \nhgal), (2) intrinsic column density (free),
(3) source redshift (fixed at $z_{em}$), (4) spectral slope (fixed at
$\aE=1.5$), and (5) model normalization (fixed at the PIMMS
normalization).

Table\,2 shows the aperture-corrected PSPC count rate for each QSO,
the derived X-ray luminosity, and \aox.  We then list
the total column density at $z=0$ required to obtain that count rate
assuming the typical RQ QSO value of $\aox= 1.6$.  The next column
shows the intrinsic column density, \nhintr, required if the absorber is
at the source redshift, with formal $1\sigma$ errors from XSPEC.
We find $\aox>1.8$ and $\nhintr >2\times 10^{22}$\,cm\mtwo\, for all
the {\em bona fide} BALQSOs with sensitive X-ray observations
(but see notes on PG1416$-$129 and 0312-555 in section\,\ref{indiv}).

We also tested our technique on a sample of 10 RQ QSOs studied by Laor
et al. (1994).  We find a sample mean of $\aox= 1.45\pm 0.08$.
To obtain the observed \ros\, PSPC count rates, none of
the 10 QSOs required additional ($z=0$) column more than
$1\times10^{20}$ cm\mtwo over the Galactic value.  Indeed, 8 of the
QSOs had higher PSPC count rates than we predicted, showing that
our choice of $\aox=1.6$ results in conservative estimates of the
absorption.

All our calculations assume $\alpha_E=1.5$, and a cold absorber at solar
metallicity.  In the following section, we discuss the effects
of these assumptions.

\section{Estimating the Intrinsic Absorbing Column \nhintr}
\label{assume}

Since X-ray spectra cannot be adequately modeled for the weak or
undetected QSOs in our sample, we must make several assumptions in
order to derive an estimate of the intrinsic column.  Essentially, we
model the expected X-ray flux $f_x$ as a function
$f(B,\nhgal,\aox,\aE,Z,U,\nhintr)$, where $Z$ is the metallicity
expressed in solar units, and $U$ is the ionization parameter.  In
most cases here (the non-detections), we know only $B$ and \nhgal, and
we assume values for the next 4 parameters (\aox, \aE, $Z$, and $U$)
to derive \nhintr.  In the discussion following, we briefly examine
the effects of these 4 parameters on our results.

\subsection{Optical-to-X-ray Slope \aox}

Our assumed value of \aox\, is 1.6, which we take to represent the
intrinsic spectral energy distribution (SED) of the QSO, i.e., before
absorption by BAL material.
Since there may already be some absorption in normal RQ QSOs affecting
this value, we are in effect deriving the additional absorption above
a typical RQ QSO, rather than relative to the bare continuum.
However, the difference is probably not important, since the majority
of optically-selected QSOs observed in soft X-rays to date show little
evidence for any substantial \nhintr\, (e.g., Fiore et al.  1994,
Bechtold et al. 1994, Walter \& Fink 1993).

Smaller assumed values of \aox\, would increase our estimated
values of \nhintr.  For RQ QSOs, a range of values from about 1 to 2
is observed (Green et al. 1995).  The difference between our assumed
value of 1.6 and the extremes of this range represents a factor of up
to 30
in our predicted broadband fluxes. The lower-than-expected soft X-ray
fluxes of BALQSOs could be entirely accounted for without excess
absorption if the intrinsic SEDs of BAL QSOs have \aox$\gapprox$1.8.
However, given such different SEDs, we would expect much larger
differences between the continua and emission lines of BAL and non-BAL
QSOs than are seen (WMFH).  On the other hand, there is as yet no
detailed study of whether large differences in UV spectral properties
are indeed seen in QSOs with widely divergent \aox.  Such a study is
now underway using $IUE$ and LBQS spectra (Green et al. 1996).  The
bottom line is that our assumed value of \aox\, probably results in
conservative estimates of the intrinsic absorption, as determined
above from tests on a similarly-sized sample of radio quiet
non-BALQSOs from Laor et al. (1994).

\subsection{X-ray Spectral Slope \aE}

If the intrinsic X-ray spectra were not as steep as we assume
(i.e., if $\alpha_E<1.5$),
then our estimates of \nhintr\, would be systematically lower.  Fits to
high signal-to-noise X-ray spectra of low redshift QSOs yield values of
\aE\, from $\sim$1.0 to 2.2 about the mean of $\aE\sim1.5$ that we use here
(Laor et al. 1994, Fiore et al. 1994).  Such a reasonable variation of
the assumed spectral slope in the X-ray bandpass makes no more than a
20\% difference in the derived broadband flux, and correspondingly no
more than a $\sim30\%$ difference in the derived \nhintr.

 An intrinsic strong soft excess, or spectral curvature, typically
below rest-frame energy 0.5\,keV, may exist in a large fraction of RQ
QSOs (Schartel et al. 1995).  Our
analysis is not affected for all but the 3 lowest redshift QSOs in our
sample, since the PSPC bandpass excludes this
spectral region.  For these 3 QSOs, our assumption of a simple power-law of
$\aE=1.5$ results in a {\em more conservative} lower limit to \nhintr.

\subsection{Ionization Parameter $U$}

We have assumed a cold, neutral gas as the intrinsic absorber.  The
soft X-ray opacity of the absorbing gas decreases with increasing
ionization state, so that our assumption results in conservative
estimates of the actual gas column.  If instead of neutral gas, we
assume $U\sim 0.1$, based on photoionization models of the OUV data
(Hamman, Korista, \& Morris 1993; HKM hereafter), then our typical
result of a few times $10^{22}$\,cm\mtwo\, would rise by about an order
of magnitude.  Even higher ionization parameters are likely to be
required to reproduce observed spectra; $U\sim
0.1$ was derived assuming lower column  densities than we find here.
Newer theoretical calculations with column densities $N_H$$\gapprox$
$10^{22}$\,cm\mtwo\, require  1$\lapprox$$U$$\lapprox$60 (Murray et al.
1995, Hamann et al. 1995).

\subsection{Metallicity $Z$}

We have assumed solar metallicity. Emission line ratios in high
redshift QSOs suggest higher $Z$ values in non-BALQSOs (e.g., Hamman
\& Ferland 1993).  Studies of the chemical enrichment of gas
in BALQSOs (e.g., HKM)
support values between 2 and 10 solar.  Our estimated values of
\nhintr\, are inversely proportional to the assumed metallicity.
Unless the nuclear
environment typically has abundances 100 times solar (Turnshek 1988),
we still require column densities at least an order of magnitude
higher than typical for RQ QSOs (Fiore et al. 1994), all else being equal.

\section{Notes on Individual Objects}
\label{indiv}

%\subsection{0043+008 = UM\,275}
%\subsection{0059-275 = LBQS}
%\subsection{0226-104 = MacAlpine}
%This QSO has a double-trough CIV BAL system (Korista et al. 1993).

\subsection{0312-555 = MZZ\,9592}
At $B=21.8$, this is the optically faintest QSO in our sample.  Even
with a total PSPC exposure time of 56\,ksec, the X-ray observation is
not sufficiently sensitive to put interesting limits on \aox\, or
\nhintr\, for this object. An exposure time of at least $\sim$250\,ksec
would be required.

%\subsection{0340-450 }
% Not serendipitous probably, but why 3.7' off-axis?

\subsection{0759+651 = IRAS 07598+6508}

This object is a member of the rare class of low-ionization BALQSOs.
These show a strong blend of broad absorption and emission lines in
both low and high states of ionization.  It is also an extreme
(optical) Fe\,II emitter, and an `ultraluminous' IRAS source.  Hines
and Wills (1995) find the spectral energy distribution to be
reddened by $E(B-V)\sim0.12$ relative to non-BALQSOs.  Sprayberry \&
Foltz (1992) find that such reddening is typical for low-ionization
BALQSOs, and accounts for the spectral differences from the generally
bluer high-ionization BALQSOs.

Assuming a gas-to-dust ratio given by Burstein \& Heiles (1978), and
the same reddening equation used in our calculation of $L_{opt}$, the
observed OUV reddening implies an equivalent Galactic column of
only $\sim10^{21}$ cm\mtwo. From our X-ray analysis, we find a ($z=0$)
column $>2.0\times 10^{22}$ cm\mtwo.
This illustrates that the absorber is probably warm -- sufficiently
ionized that its effects are not seen in the OUV, and with a
low dust-to-gas ratio.  A low dust-to-gas ratio might seem strange in an
IRAS QSO where much of the far-IR luminosity is expected to be from
warm dust (possibly merger related).  However, much of the IR-emitting
region may be extranuclear.

%\subsection{0932+501 }

\subsection{1246-057 }
Based on the PSPC image, there appear to be two sources near to the
optical QSO position.  Source A is at $12^h49^m13.1\,
-05^{\circ}59'13.0''$, only 14$''$ from the optical QSO position.
Source B, at $12^h49^m14.7\,-06^{\circ}00'34.6''$, is 77$''$ from the
optical QSO position and has no obvious optical counterpart on the
POSS.  We assume that source A is the BALQSO.  Net counts in the hard
(0.41 - 2.45\,keV) band are $74\pm 12$. The soft (0.12 - 0.40\,keV) band
shows only a $3\sigma$ upper limit of 32.4 counts.

Although this is one of only 2 detections (see PG1416$-$129 below)
in our sample, it still shows a value of \aox\, that is at the extreme
X-ray weak end of the observed range for QSOs.  The
intrinsic column we derive, $1.2\times 10^{23}$ cm\mtwo, is the highest
in the sample.  This indicates that our lower limits to \nhintr,
in the case of non-detections, are probably conservative.

%\subsection{1309-056 }
%\subsection{1413+117 }

\subsection{1416$-$129 = PG }

 PG1416$-$129 is a strong detection that has been analyzed by deKool
\& Meurs 1994. Their results for the ROSAT bandpass were quite typical
for a RQ QSO ($\alpha_E=1.2$), while earlier GINGA results were very
flat ($\alpha_E\approx 0.1$, Williams et al. 1992).  The 4 year
separation of those observations, however, could render the data difficult
to interpret, since flat-spectrum QSOs in particular can vary
enormously.  The BAL classification was originally awarded by Turnshek
\& Grillmair (1986), based on a single $IUE$ spectrum in each
of the $SWP$ and $LWR$ instruments.  However, we have examined the
higher quality, optimally-extracted and co-added spectrum of
PG1416$-$129 of Lanzetta, Turnshek, \& Sandoval (1993), and we believe
that the object is {\em not} a BALQSO.  At the very least, it
unquestionably shows the weakest UV absorption of any QSO in our
sample.  An $HST$ spectrum would resolve the issue once and for all.

Using the standard technique we have described (which assumes
$\aE=1.5$ and only \nhgal), we derive an expected count rate
consistent with that observed.  That is, no \nhintr\, is required.
Furthermore, \aox\, for this QSO is on the {\em low} (X-ray bright) end
of the range for RQ QSOs.  These X-ray results, consistent with
deKool \& Meurs (1994) thus also suggest that PG1416$-$129 is not a
{\em bona fide} BALQSO.

%\subsection{1700+518 = PG}
%PG1700+518 is a low redshift ($z=0.288$) BALQSO.
%Morphological structure investigated by Hutchings et al. (1992).

\section{Conclusions}
\label{conclude}

Using deep pointed observations from the \ros\, PSPC, we demonstrate
that BALQSOs are weak in the soft X-ray bandpass in comparison to RQ
QSOs with normal OUV spectra.  The only object in our sample with a
sensitive X-ray observation that is not anomalously weak in the \ros\,
PSPC bandpass is PG$1416\!-\!129$, which has a low value of \aox\, and
thus requires no intrinsic absorption.  However, of all the QSOs in
our sample, PG$1416\!-\!129$ has the weakest BALs.  Consideration of
optimally-extracted $IUE$ data and our ROSAT analysis lead us to
propose that it is not a {\em bona fide} BALQSO.

 True BALQSOs have proved notoriously difficult to detect in X-rays,
but evidence until recently has been mostly anecdotal.  As an example,
Bregman (1984) obtained \ein\, IPC observations of 3 BALQSOs, and
reported detection of only one, UM\,232, which showed no evidence of an
absorption cutoff.  It turns out that the X-ray source is not UM\,232,
but a low-redshift, non-BAL QSO projected nearby (Junkkarinen, Barlow,
\& Cohen 1993).  Our work extends and complements that of Green et al.
(1995).  Their sample of 36 BALQSOs was chosen from a large
uniformly-selected QSO sample (the LBQS), and forms the most
`complete' sample ever observed in X-rays.  The short ($\sim 600$\,sec)
exposure times of the \ros\, All Sky Survey meant that the upper limits
(for 35 of the 36 QSOs) were not very sensitive.  By stacking the
X-ray data, they achieved a total exposure time of about 21\,ksec. By
using the mean
\nhgal and redshift for that sample, we find here that no intrinsic
column is required within the large errors.  However, Green et al.
(1995) were able to show that the BALQSO sample was X-ray quiet at the
99.5\% level compared to carefully chosen comparison RQ QSO samples.
Although the sample we study here is by nature much more
heterogeneous, the mean exposure time of $\sim25$\,ksec for {\em each}
object means that we are able to provide much more interesting limits
($\aox>1.8$) on the optical-to-X-ray slope of
{\em bona fide} BALQSOs with sensitive X-ray observations.

We speculate that large values of \aox\, may prove to be a defining
characteristic of BALQSOs.  At the very least, we predict that samples
of QSO candidates with large \aox\, will yield a higher percentage of
BALQSOs, particularly at low redshift, where BALs are less easily
detected in the optical, and X-ray surveys are more sensitive to low
luminosity objects. As a corollary, X-ray-selected QSO samples should
yield fewer BALQSOs.

Several arguments in the literature suggest that BALQSOs are not a
separate population $\sim10\%$ as large as normal RQ QSOs, but rather
RQ QSOs viewed along a line of sight through an absorber typified by a
$\sim10\%$ covering fraction.  First, the emission line parameters
indicate similar ionizing continua in both BAL and non-BAL QSOs
(WMFH).  Also, constraints on the covering fraction of BALQSOs
(HKM) suggest that all RQ QSOs have BAL regions.  In any case, the
intrinsic spectral energy distributions of
BALQSOs are likely to be similar to their unabsorbed analogs, but
BALQSOs provide a special view into the gas dynamics around normal RQ
QSOs.  Although with these data, we cannot definitively identify the
OUV and X-ray absorbers as the same material, based on the arguments just
presented, we expect the large observed values of \aox\, in BALQSOs
may be caused by absorption by high column density, ionized,
outflowing material along the line of sight.  If this is indeed the
main difference between BALQSOs and normal RQ QSOs, the intrinsic
absorbing columns in BALQSOs are more than 100 times higher.  Using a
sample of 12 QSOs observed in the soft X-ray regime by the ROSAT PSPC,
we place high lower limits on this absorption of
$\nhintr >2\times 10^{22}$\,cm\mtwo\, for 10 QSOs.  Of the two that do
not require excess absorption, one is at best a marginal BALQSO
(1416-129), while the other (0312-555) has an X-ray observation too
shallow to provide sensitive limits to \nhintr.  The one detected,
definitive BALQSO, $1246\!-\!057$, yields the highest column estimate
of all, $\nhintr=1.2\times 10^{23}$ cm\mtwo.

We note that some variability has been seen in the BAL spectra of 3
QSOs in our sample.  Smith \& Penston (1988) found changes in the BALs
of 1246$-$057 and possibly 1309$-$056 at the $\sim20\%$ level,
indicating lower limits to column changes of only about
$10^{14}$cm\mtwo.  A demonstration of correlated variability between
BALs and soft X-ray flux would provide more evidence that the UV and
soft X-ray absorbers are physically associated, but is a daunting task
given the weak X-ray fluxes.  The case for 1413+117 is more
complicated, since it is gravitationally lensed (Hutsemekers 1993).

Most models for BALQSOs (e.g., Arav, Li, \& Begelman 1994) have been
tailored to reproduce the columns inferred from OUV data alone. At
least one recent model (Murray et al. 1995) predicts a column density
of $10^{23} - 10^{24}$ cm\mtwo, consistent with the total absorbing
columns we derive here.  Murray et al. (1995) explain BAL spectra in
QSOs as the result of a (nearly edge-on) line of sight along an
accretion disk, where the inner edge of a radiatively accelerated wind
entrains dense gas off the disk.  The large column and warm ionization
state of this gas blocks the soft X-rays from the nucleus, but
transmits UV photons.  In this way, cloud material further out can
continue to be radiatively accelerated to $\sim0.1c$, since its CNO
ions will not be entirely stripped by the soft X-ray emission.  This
model is consistent with our findings here of very high intrinsic column
densities in BALQSOs.

The much lower absorbing columns (e.g., $N_H\sim 10^{19}$ to
$10^{20}$) inferred for the BAL clouds from the OUV data (HKM,
Turnshek 1988) are such that {\it a priori} we would have expected
very little soft X-ray absorption ($\tau<<1$).  Soft X-ray
observations thus prove useful in constraining the total line-of-sight
column.  If the BALQSO phenomenon is indeed only a viewing angle
effect, then in the hard X-ray range ($>$3\,keV), BALQSOs should have
spectral slopes similar to normal radio quiet QSOs.  Some early
evidence for this is that an ASCA spectrum of the prototype BALQSO,
PHL\,5200, is best fit using a typical RQ QSO power-law with large
intrinsic absorption, \nhintr a few times $10^{23}$ (Mathur, Elvis, \&
Singh 1995).  We have been awarded XTE time to further pursue hard
X-ray studies of BALQSOs.

Many thanks to Tom Barlow for providing us a list of BALQSOs with his
valuable comments.  Belinda Wilkes' comments substantially
improved our first draft.  Support for PJG was provided by the NSF
through grant INT\,9201412, and by NASA through Grant HF-1032.01-92A
awarded by the Space Telescope Science Institute, which is operated by
the Association of Universities for Research in Astronomy, Inc., under
NASA contract NAS5-26555. SM was supported by NASA grant NAGW-4490
(LTSA).  This research has made use of data obtained
through the High Energy Astrophysics Science Archive Research Center
Online Service, provided by the NASA-Goddard Space Flight Center,
and of the Simbad database, operated at CDS, Strasbourg, France.

\clearpage

\hoffset-0.5in

\begin{centering}
\begin{tabular}{cccclccrrrrrc}
\multicolumn{13}{c}{ TABLE 1 } \\
\multicolumn{13}{c}{ RAW DATA FOR BAL QSOS WITH ROSAT PSPC POINTINGS } \\
\hline\hline
\multicolumn{13}{c}{ } \\
 & & & $N_{H}^{Gal}\,^c$ & PSPC & Off-Axis & \multicolumn{1}{c}{Exposure$^d$} &
\multicolumn{4}{c}{Counts} & \multicolumn{1}{c}{$R_{ap}$} & $B$ mag \\
 Name &  $B^a$  &  z$^b$  & ($10^{20}$cm\mtwo) & Sequence & Dist ($'$) &
\multicolumn{1}{c}{(s)} & Source & \multicolumn{1}{c}{$\sigma$} & Bkg
&  \multicolumn{1}{c}{$\sigma$}  & \multicolumn{1}{c}{($''$)} & Refs \\ \hline
0043+008  & 17.2 &  2.141 & 2.03 & rp700377 & 0.3 & 10858 & $<34.0$ & \ldots &
79.8 & 2.8 & 90 & 1 \\
0059$-$275 & 18.0 & 1.594 & 1.99 & rp700121AO2 & 0.1 & 15154 & $<36.6$  &
\ldots & 104.4 & 3.2 & 90 & 1 \\
\multicolumn{4}{c}{ }  & rp700121 & 0.1 &  \multicolumn{7}{c}{ }  \\
0226$-$104 & 17.0 & 2.256 & 2.59 & rp700114 & 0.4 &  9517 & $<36.0$ & \ldots &
76.0 & 3.4 & 90 & 1 \\
0312$-$555 & 21.8 & 2.710 & 1.68 & wp701036N00 & 11.5 &  56213 & $<54.0$ &
\ldots & 299.5 & 3.8 & 75 & 2 \\
\multicolumn{4}{c}{ }  & wp701036AO1 & 11.5 & \multicolumn{7}{c}{ }  \\
0340$-$450 & 18.7 & 2.004 & 1.61 & rp900495N00 & 3.7 &  74784 & $<60.0$ &
\ldots & 367.40 & 4.4 & 75 & 3 \\
\multicolumn{4}{c}{ }  & rp150029 & 3.7 & \multicolumn{7}{c}{ }  \\
0759+651 & 14.5 & 0.148 & 4.33 & wp700258 & 0.1 & 8354 & $<32.4$ & \ldots &
74.0 & 1.6 & 90 & 4 \\
0932+501 & 17.1 & 1.914 & 1.35 & rp400243 &10.7 & 10184 & $<24.0$ & \ldots &
53.6 & 1.7 & 90 & 1 \\
\multicolumn{4}{c}{ }  & rp400244 &9.8 & \multicolumn{7}{c}{ }  \\
\multicolumn{4}{c}{ }  & rp400245 & 9.8 & \multicolumn{7}{c}{ }  \\
1246$-$057 & 17.1 & 2.222 & 2.10 & rp60026 & 14.7 & 57591 & 84.4 & 15.7 & 690.1
& 8.8 & 45 & 1 \\
\multicolumn{4}{c}{ }  & rp60026AO1 & 14.7 &  \multicolumn{7}{c}{ }  \\
\multicolumn{4}{c}{ }  & rp60026AO2  & 14.7 & \multicolumn{7}{c}{ }  \\
1309$-$056 & 17.2 & 2.212 & 2.56 & rp700376 & 11.8 & 12197 & $<34.9$ & \ldots &
89.2 & 2.4 & 75 & 1 \\
1413+117 & 17.0 & 2.542 & 1.78 & rp700122 & 0.1 & 28066 & $<80.1$ & \ldots &
576.7 & 8.0 & 120 & 5 \\
1416$-$129 & 15.7 & 0.129 & 7.20 & wp700527 &0.3 &  9109 & 3440. & 61. & 177.7
& 8.0 & 120 & 5 \\
1700+518 & 15.1 & 0.288 & 2.67 & rp700123 &0.2 &  8196 & $<26.1$ & \ldots &
65.1 & 2.2 & 90 & 5 \\
\hline
\end{tabular}\\[0.5ex]

\hskip-0.2in \parbox[b]{7in}{
\,\,\,$^a$\,$B$ magnitude includes BAL $k$-correction from Stocke et al.
(1992).

\,\,\,$^b$\,Redshifts from Korista et al. (1993) when available, otherwise
{}from Hewitt \& Burbidge (1993).

\,\,\,$^c$\,\nhgal \,from Stark et al. (1992) except for 0043$+$008 and
1416$-$129 from Elvis, Lockman, \& Wilkes (1989).

\,\,\,$^d$\,Summed exposure time of all listed PSPC sequences.}
 \hskip-0.2in \parbox[b]{7in}{
\,\,\,REFERENCES.---(1) Stocke et al. 1992
(2) Marano, Zamorani, \& Zitelli 1988;
(3) Boyle et al. 1980;
(4) Low et al. 1989;
(5) Schmidt \& Green 1986
}

\end{centering}

\clearpage
\vskip2cm
\begin{centering}
%\begin{table}[thpb]
\begin{tabular}{lrrrrrr}
\multicolumn{7}{c}{ TABLE 2 } \\
\multicolumn{7}{c}{ DERIVED PROPERTIES OF BAL QSOS WITH ROSAT PSPC
POINTINGS } \\
\hline\hline
\multicolumn{7}{c}{ } \\
   & \multicolumn{1}{c}{Count Rate} & & &
\multicolumn{3}{c}{ $N_H^b$}  \\
  Name  & \multicolumn{1}{c}{(corrected)} &$l_{2{\rm keV}}^a$ & \aox
& ($z=0$) &  ($z=z_{em}$) & error \\  \hline
$0043\!+\!008$ & $<\!$0.0034 & $<\!$27.00 & $>\!$1.89 & 35 &  560 & 230 \\
$0059\!-\!275$ & $<\!$0.0027 & $<\!$26.55 & $>\!$1.83 & 20 &  180 & 100 \\
$0226\!-\!104$ & $<\!$0.0041 & $<\!$27.24 & $>\!$1.83 & 30 &  500 & 220 \\
$0312\!-\!555^c$ & $<\!$0.0019 & $<\!$26.70 & $>\!$1.36 & \ldots & \ldots &
\ldots \\
$0340\!-\!450$ & $<\!$0.0009 & $<\!$26.27 & $>\!$1.90 & 30 &  420 & 140 \\
$0759\!+\!651$ & $<\!$0.0042 & $<\!$24.66 & $>\!$2.13 & 200 &  280 & 50 \\
$0932\!+\!501$ & $<\!$0.0025 & $<\!$26.62 & $>\!$2.00 & 55 &  790 & 380 \\
$1246\!-\!057$ &  0.0019 &  26.62 &  1.98 & 65 &  1240 & 130 \\
$1309\!-\!056$ & $<\!$0.0032 & $<\!$27.09 & $>\!$1.84 & 30 &  480 & 200 \\
$1416\!-\!129^c$ & 0.3863    &  26.56     & 1.40  & \ldots & \ldots & \ldots \\
$1413\!+\!117$ & $<\!$0.0029 & $<\!$27.08 & $>\!$1.93 & 40 &  920 & 160 \\
$1700\!+\!518$ & $<\!$0.0034 & $<\!$25.02 & $>\!$2.30 & 250 & 460 & 80 \\
\hline
\end{tabular}\\[0.5ex]

\hskip0.5in \parbox[b]{5in}{
\,\,\,$^a$Monochromatic luminosity at 2keV in \lnucgs.

\,\,\,$^b$Inferred intrinsic column density in units $10^{20}$cm$^{-2}$.

\,\,\,$^c$See discussion in \S5.

}

\end{centering}
\end{document}